# On the modular platoon-based vehicle-to-vehicle electric charging problem

**Zhexi Fu, Joseph Y. J. Chow**[*]
C2SMART University Transportation Center, NYU Tandon School of Engineering, Brooklyn, NY
*Corresponding author: joseph.chow@nyu.edu

**Abstract**

We formulate a mixed integer linear program (MILP) for a platoon-based vehicle-to-vehicle charging (PV2VC) technology designed for modular vehicles (MVs) and solve it with a genetic algorithm (GA). A set of numerical experiments with five scenarios are tested and the computational performance between the commercial software applied to the MILP model and the proposed GA are compared on a modified Sioux Falls network. By comparison with the optimal benchmark scenario, the results show that the PV2VC technology can save up to 11.07% in energy consumption, 11.65% in travel time, and 11.26% in total cost. For the PV2VC operational scenario, it would be more beneficial for long-distance vehicle routes with low initial state of charge, sparse charging facilities, and where travel time is perceived to be higher than energy consumption costs.

# 1. Introduction

The U.S. transportation sector contributed to the largest share (28%) of greenhouse gas (GHG) emissions in 2021, of which over 94% primarily came from burning petroleum-based fossil fuels including gasoline and diesel (EPA, 2024). In Europe, the European Parliament (EP) banned the sales of new vehicles with internal combustion engines (ICEs) from 2035 to reach a carbon-neutral goal by 2050 (EP, 2024). However, long charging times hamper mass EV adoption (Jung et al., 2014; Jung & Chow, 2019). On average, it takes 20-30 minutes for DC fast chargers to fill up an 80-mile battery, and 4-6 hours for Level 2 chargers (Chargepoint, 2020; Liu et al., 2023). Furthermore, the charging rate is nonlinear and becomes much slower above an 80% state of charge, depending on factors such as the real-time terminal voltage and the number of EVs getting charged simultaneously (Montoya et al., 2017). As a result, drivers need to plan for detours to charging stations (CSs) and allow for charging delays due to the limited range and limited availability of charging facilities, especially for long-distance trips.

Aside from increasing the range of EVs through innovative battery design and advanced battery management systems, many efforts have been studied to improve EV charging delivery mechanisms, which can be divided into two major categories: stationary and dynamic. One stationary option is to simply construct a dense enough charging station network. However, unbalanced charging demand patterns might cause long charging times and delays. A second stationary option is to deploy battery swapping stations along with battery storage to hold fully charged batteries that are swapped with vehicles' depleted batteries. A third stationary option is deploying wireless charging lanes or electrified roads which can recharge vehicles without them having to come to a full stop (Jeong et al., 2015; Chen et al., 2016; Schwerdfeger et al., 2021; Tran et al., 2022). However, the costs can be quite high, and difficulty in construction, maintenance. and future upgrades might limit the actual performance once implemented.

Dynamic options consider charging delivery that is not fixed to a static location. One such solution is to deploy on-demand mobile chargers (MCs) or battery swapping to charge EVs on-site, defined as charging as a service (CaaS). This solution mitigates the queue delays during peak hours at the CS, but may suffer from operating costs of delivering the batteries or charge (Abdolmaleki et al., 2019; Qiu and Du, 2023). The latter battery swapping solution is also called Battery-as-a-Service (BaaS). However, the overall performance might be limited by the number of back-up batteries at each station and the compatibility design of EVs, not to mention other new problems, such as battery ownership and innovation inhibitions. According to NIO (2024), their latest second-generation battery swap station can only store up to 14 batteries and serve up to 312 EVs per day.

Thanks to the recent development of vehicle-to-vehicle (V2V) charging technology (Mou et al., 2020; Rostami-Shahrbabaki et al., 2022) along with modular vehicles (MVs) that can couple and decouple on the move (Guo et al., 2018; Caros and Chow, 2021; Dakic et al., 2021; Fu and Chow, 2023), an innovative dynamic charging solution involving platoon-based vehicle-to-vehicle charging (PV2VC) may be feasible. Since MVs can travel in a platoon with a physical connection, electrified MVs can be deployed as electricity suppliers (ESs) to transfer power to other EVs while moving together to avoid the detour and delay at a CS. As shown in **Figure 1**, an electricity request (ER) with a low battery level can be charged while moving together with an ES in a platoon. This type of dynamic charging solution has attracted research attention lately (Maglaras et al., 2014; Kosmanos et al., 2018; Abdolmaleki et al., 2019; Liu et al., 2019; Chakraborty et al., 2022; Qiu and Du, 2023). However, the literature does not yet consider the vehicle routing problem that either serve passengers or provide charge to other vehicles.

To evaluate the operational feasibility of PV2VC technology in that manner, we break down the technology into three scenarios of increasing complexity. The first scenario, representing the best practice benchmark, is the electric vehicle routing problem (EVRP). In this case, electric vehicles get charged at stationary locations, such as charging stations and battery swap stations. In the second scenario, vehicles still charge at CSs, but can form platoons along their travel path to reduce their energy consumption, defined as the electric vehicle platooning problem (EVPP). Only a handful of studies have touched on the variant EVPP so far.

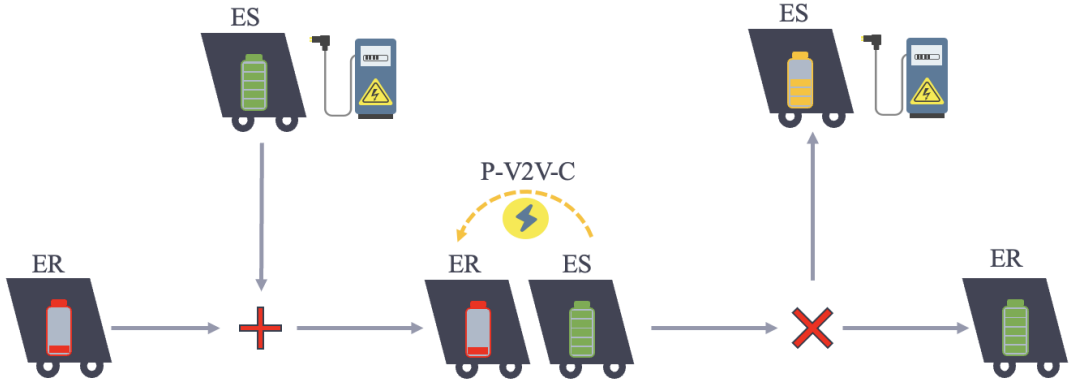
Figure 1. Demonstration of PV2VC technology

The third scenario is the PV2VC problem, which has not yet been formulated or solved. Since ERs and ESs need to travel in an attached platoon to transfer electric power, tracking the movement and status of EVs and temporal-spatial synchronization is required to ensure the precision and efficiency of the entire operation, especially the platoon formation and separation processes. In addition, we aim to jointly optimize both the routing and charging scheduling to minimize the energy consumption and travel time of ERs.

Overall, this study makes the following major contributions. First, we introduce a planning model for a novel dynamic charging solution to address the detour and charging delays of EVs, defined as a static PV2VC problem. ERs can travel in an attached platoon to reduce energy consumption and receive electricity transfer from ESs while moving toward their destinations. Second, we mathematically formulate the PV2VC problem to minimize the total energy consumption and travel time of ERs in a mixed integer linear programming (MILP) model, along with the two problem variants of EVRP and EVPP. Third, a customized genetic algorithm (GA) is proposed to solve the large-scale PV2VC problem for realistic instances.

The paper is organized as follows. Section 2 first demonstrates the research problem through an illustrative example of the design problem followed by a review of prior work. Section 3 presents the proposed MILP model. The proposed GA is shown in Section 4. We conduct numerical experiments in Section 5. Section 6 provides a conclusion with future directions.

## 2. Literature review
### 2.1 Problem illustration
We consider three operational scenarios: 1) a single vehicle that can only get charged at a charging station (CS); 2) based on the first scenario, vehicles are allowed to platoon to save energy; and 3) the PV2VC technology where vehicles can charge on the move through power transfer between connected modular vehicles. The objective is to minimize the total energy consumption and travel time for electricity requests (ERs), including the charging time at a CS, platoon formation delay, and wait time for the PV2VC service.

Consider two electricity requests (ERs: $r_1$ and $r_2$) and one electricity supplier (ES: $s_1$) on a 5-node directed graph shown in **Figure 2**. Each node is an initial vehicle location, a task for ER to visit, a potential location for joining or splitting a platoon, or a charging station for recharging ER and ES. Both ERs are initially located at node 0, where $r_1$ has a battery level of 20 kWh and $r_2$ has 35 kWh. The $r_1$ has a sequence of tasks to visit at nodes [1,3] and $r_2$ has a sequence of tasks to visit at nodes [3,4]. ERs are allowed to detour for recharging at a CS, or travel in platoon with others to save cost in between any two tasks (noting in practice, ERs should not be serving customers if they need to recharge in the middle of service). The ES $s_1$ is initially located at node [2] with a battery level of 45 kWh. The travel distance is shown next to each link in **Figure 2** and two charging stations are located at nodes [2, 4].

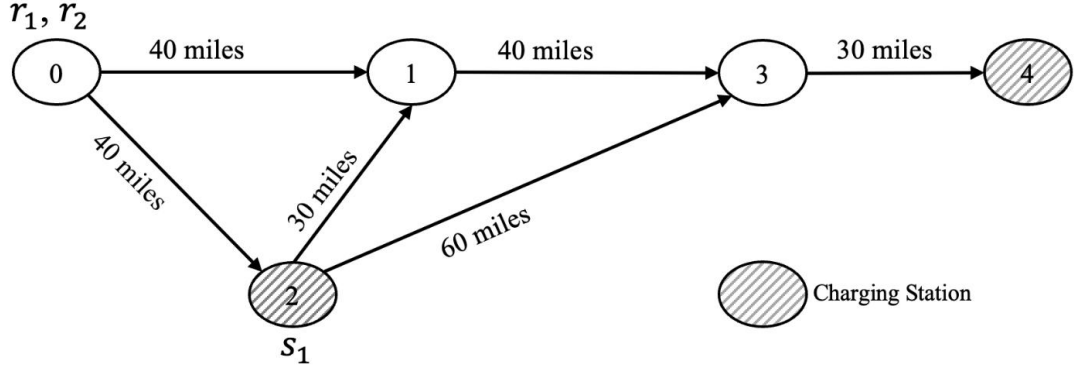

Figure 2. Illustrative example

In this example, we make the following assumptions. The ER has a battery capacity of 90 kWh and the ES has double the capacity at 180 kWh. The charging rate at the CS is assumed to be 180 kW and 50 kW for vehicle-to-vehicle charging with a transfer efficiency as of $\eta_2 = 0.9$ (the ER receives 90% of the output power transferred from ES, which is 45 kWh here). We assume that an ER must keep a minimum safety battery level of 2 kWh at any node and an ES can only move to a node (including providing the PV2VC service en-route) if the remaining battery level after arriving at this node is still enough to reach the nearest CS. This setting is intended for electric vehicle relocation and prevents running out of the vehicle battery (Pantelidis et al., 2022).

Similar to Abdolmaleki et al. (2019) and Qiu and Du (2023), one ES can only charge one ER over a link at the same time (constant output/input power for ES/ER, respectively). This is to avoid the scenario where multiple ERs receive energy from multiple ESs, which might require additional research on optimal distribution of electricity among ERs, the transmission efficiency across different positions in a platoon, and power transfer upper/lower thresholds.

Unlike those two studies, we do not limit the platoon size, and multiple ERs are allowed to travel in the same platoon at the same time. If one ER is matched and served by an ES while traveling in a platoon, we then consider that the associated ES is also traveling in the same platoon. Thus, for a platoon consisting of $n$ ERs, there could be 0 to $n$ ESs in the platoon as well, which leads to a maximum platoon size of $2n$. Furthermore, we assume that all vehicles travel at a constant speed of 60 mph and the electricity consumption rate is 0.4 kWh/mile. Vehicles traveling in a platoon receive a constant $\eta_1 = 10\%$ savings in energy cost. All ERs and ESs are available at time $T = 0$ and that join/split times are negligible without loss of generality.

**Table 1** summarizes the optimal routes, charging schedules, and various costs for the illustrative example over three scenarios. In the baseline scenario, both $r_1$ and $r_2$ detour to node {2} to get charged before heading to finish their tasks. This solution leads to energy costs of 44 kWh and 52 kWh, charging times of 8.67 min and 6.33 min, and travel times of 110 min and 130 min for $r_1$ and $r_2$, respectively. The total cost is 351 min, assuming the cost of 1 kWh is equivalent to 1 minute of time cost.

In the second operation scenario where vehicles can travel in a platoon to save energy consumption, the optimal routes for $r_1$ and $r_2$ are to travel in a platoon through node 0 to the CS at node 2. After charging for 8.13 and 5.8 mins, $r_1$ and $r_2$ continue to their remaining tasks separately. In this case, the travel time stays the same as the first scenario and the total cost equals 346.73. Due to the energy savings from platooning, $r_1$ and $r_2$ can save 3.64% and 3.08% from their electricity consumption and 6.15% and 8.42% from their charging times at a CS, which leads to 1.22% savings in total cost.

Table 1. Summary of three operation scenarios

| Operation Scenario | Vehicle | Optimal Route | Energy Cost (kWh) | Charging Time at CS (mins) | Travel Time (mins) | Total (min) |
|---|---|---|---|---|---|---|
| | $r_1$ | 0→2→1→3 | 44 | 8.67 | 110 | 351 |
| **Single vehicle with CS** | $r_2$ | 0→2→3→4 | 52 | 6.33 | 130 | |
| **Platoon with CS** | $r_1$ | **0**→**2**→1→3 | 42.4 (-3.64%) | 8.13 (-6.15%) | 110 | 346.73 (-1.22%) |
| | $r_2$ | **0**→**2**→3→4 | 50.4 (-3.08%) | 5.8 (-8.42%) | 130 | |
| **PV2VC** | $r_1$ | **0**→**1**→**3** | 28.8 (-34.55%) | 0 | 80 (-27.27%) | 263.05 (-26.38%) |
| | $r_2$ | **0**→**1**→**3**→4 | 39.6 (-23.85%) | 0 | 110 (-15.38%) | |
| | $s_1$ | 2→**1**→**3**→**4** | 37.2 | 3.84 | 100 | 141.04 |

Notes: 1) vehicle routes with underline and bold stand for platooned segments.
 2) percentages in "()" are cost differences of the second/third scenario ($Y$) against the first scenario ($X$), calculated as $\frac{Y-X}{X} \times 100\%$, where a negative sign represents lower cost and savings.

In the PV2VC scenario, ERs can charge while they are moving towards the tasks without detouring to the CS, so the optimal routes for $r_1$ and $r_2$ are to travel in a platoon through nodes [0,1,3]. The ES $s_1$ merges with the platoon of ERs at node 1 and then travel together until the last task of $r_2$ at node 4. In this scenario, both $r_1$ and $r_2$ travel in a platoon through their entire trips. The energy cost reduces to 28.8 kWh and 39.6 kWh, and the travel time is 80 min and 110 min in this case. The total cost of $r_1$ and $r_2$ drops to 263.04 min. From the perspective of ERs only, they are reduced by 34.55% and 23.85% in electricity consumption, 27.27% and 15.38% in travel time, and 26.38% in total cost across both vehicles. However, we also present the cost of employing the ES in **Table 1**. The $s_1$ gets charged at node 2 for 3.84 mins, then consumes 37.2 kWh in electricity from itself, and travels 100 mins to provide the platoon-based V2V charging services for $r_1$ and $r_2$ until node 4, which leads to a total cost of 141.04.

In this case, the total cost in the PV2VC scenario when considering both ER and ES sums to 404.09, which is a 15.13% (16.54%) more than the first (second) scenario. This illustrates the need for methods to identify circumstances when PV2VC works best and when it should not be used. The resulting savings on the ER's side is still promising (average savings of 28.75% on energy cost, 20.83% on travel time, 26.38% on total cost, and no detour for charging at CS). Furthermore, the fleet of ESs could be built on modular vehicles that can be used for other services when idle.

## 2.2 Prior research

The basic benchmark operational scenario, the EVRP, is well studied. There are studies considering time windows (Schneider et al., 2014; Hiermann et al., 2016; Desaulniers et al., 2016), capacitated charging stations (Kullman et al., 2021; Froger et al., 2022), nonlinear charging speeds (Montoya et al., 2017), partial recharges (Felipe et al., 2014; Keskin and Çatay, 2016; Cortés-Murcia et al., 2019), variant charging prices (Abdulaal et al., 2017; Lin et al., 2021) and battery swapping (Huang, et al., 2015; Raeesi and Zografos, 2020). As for solution methods, there are studies focusing on exact algorithms (Desaulniers et al., 2016; Andelmin and Bartolini, 2017), and metaheuristic approaches (Erdogan and Miller-Hooks 2012; Schneider et al., 2014; Hiermann et al., 2016; Montoya et al., 2017). For a more comprehensive literature review on the EVRP, see Kucukoglu et al. (2021).

In the EVPP, vehicles can platoon but are still required to detour to those stationary locations to get charged. The platooned vehicles can reduce energy consumption (Tsugawa et al., 2016; Song et al., 2021) and thus reduce the required en-route charging time at a CS. Only a few have studied the vehicle platooning problem (VPP) so far, not to mention the problem variant with electric vehicles. Larsson et al. (2015) developed a mixed integer linear programming model (MILP) to minimize the total energy consumption for the truck platooning problem. To solve for large problems, they presented a two-phase heuristic algorithm. Fu and Chow (2023) formulated a MILP model and proposed a modified large neighborhood search algorithm based on the Steiner tree problem.

As for the electric vehicle platooning problem (EVPP) with consideration of en-route charging at CS, Scholl et al. (2023) only considered a single unidirectional highway where trucks move in one direction. Service points are distributed on the highway where electric vehicles can get charged and form platoons. Since they only focused on a single highway, their proposed methodologies may suffer from considerable optimality gaps and even infeasible solutions if vehicles traverse multiple connected highways or just on a normal two-dimensional network. Alam and Guo (2023) formulated a MILP model to both optimize the platooning and charging strategies for long-haul electric freight vehicles. Although they conducted numerical experiments with real-world commercial freight trip data in Florida, the test is still based on a single highway route similar to Scholl et al. (2023).

Yuan et al. (2025) focused on optimizing the Modular Autonomous Electric Vehicles (MAEV) based demand responsive transit system, which is an electrified extension to Fu and Chow (2023) that involving electric vehicles and charging strategy. However, their study does not address the platoon-based vehicle-to-vehicle charging problems of electric modular vehicles.

In the scenario with modular vehicle-to-vehicle power transfer, the technology traces back to Maglaras et al. (2014), who proposed a concept of using trucks or buses that moving along national highways as mobile energy disseminators to charge surrounding electric vehicles while moving via either wired connection or electromagnetic inductive power transfer. Kosmanos et al. (2018) and Liu et al. (2019) continued the study by focusing on route optimization and vehicle assignment, respectively.

Chakraborty et al. (2022) presented a system framework with associated algorithms for the peer-to-peer car charging (P2C2) concept to enable EVs to transfer energy among each other while in motion. They also simulated the P2C2 framework on a traffic simulator, SUMO, and found that the operation cost can be reduced due to fewer number of charging stops and lower battery capacity requirements.

On an energy-space-time expanded network, Abdolmaleki et al. (2019) formulated a MIP model to optimize the matching, scheduling, and routing problem for V2V wireless power transfer (WPT) between EVs. The objective of the V2V WPT problem is to maximize the number of possible trips and minimize the total energy cost at the same time. They also developed a dynamic programming algorithm and conducted numerical experiments with trip data from Michigan.

Qiu and Du (2023) developed a MILP model for the on-the-move electric-vehicle-to-electric-vehicle charging (mE2) technology on a directed graph. Large-scale numerical experiments were implemented on citywide (Chicago) and statewide (Florida) networks, and solved by a clustering-aid decomposition and merging (c-DM) algorithm.

Both Abdolmaleki et al. (2019) and Qiu and Du (2023) only considered a maximum platoon size of two, namely one energy supplier vehicle and one energy recipient vehicle, and did not incorporate the energy saving benefits from platooning in their problem formulations. Moreover, in Qiu and Du (2023), the recipient vehicle was not allowed to detour to receive the electricity from ES (pre-defined fixed route for ER). It was only allowed to charge from the mE2 service or at charging stations (the percentage of ERs using the mE2 service or CS was pre-defined exogenously and not optimized in their numerical tests).

The key literature relevant to this study is summarized in **Table 2**.

Table 2. Summary of key literature

| Field | Study | Methodology | Major Contributions |
|---|---|---|---|
| **VPP** | Larsson et al. (2015) | MILP model, two-phase heuristic algorithm | 1) Defined the truck platooning problem in road networks with mathematical model and proved the problem as NP-hard. 2) Applied a two-phase local search heuristic algorithm for large-scale instances and showed the effectiveness of truck platooning with different start locations and vehicle numbers. |
| | Fu and Chow (2023) | MILP model, large neighborhood search algorithm | 1) Formulated the modular vehicle platooning problem on general undirected networks where these vehicles are allowed to form or leave a platoon at any location and time to minimize the energy consumption. 2) Proposed a large neighborhood search algorithm based on Steiner tree problem and demonstrated up to 29% savings in total cost against benchmark mobility on-demand system. |
| **EVPP** | Alam and Guo (2023) | MILP model | 1) Developed an optimization model to both optimize the charging and platooning for long-haul electric freight vehicles.<br>2) Compared three state-of-the-art solution algorithms on CPLEX. |
| | Scholl et al. (2023) | MIP model, adaptive large matheuristic search | 1) Derived an optimization model for electric commercial vehicles on a single highway and unidirectional travel with the consideration of recharge scheduling and trucks' willingness-to-wait for platoon scheduling.<br>2) Presented a novel matheuristic based on adaptive large neighborhood search algorithm. |
| | Yuan et al. (2025) | MILP model, adaptive large neighborhood search algorithm | 1) Proposed an optimization model to optimize the route and charging plans for modular autonomous electric vehicles (MAEV) with passenger en-route transfers.<br>2) Developed an adaptive large neighborhood search algorithm and verified the performance through large cases.<br>3) Validated the significant advantages of MAEV with en-route transfers against the standard demand-responsive transit service. |
| **PV2VC** | Abdolmaleki et al. (2019) | MIP model, dynamic programming | 1) Introduced a mathematical program to route, schedule, and match supplier and recipient vehicles on energy-time expanded network for the vehicle-to-vehicle wireless power transfer (V2V WPT) problem.<br>2) Proposed a dynamic programming solution to find the optimal vehicle routes on energy-space expanded network.<br>3) Conducted a series of numerical studies to quantify the degree of benefits of the V2V WPT technology using long-distance trip data in Michigan state. |
| | Qiu and Du (2023) | MILP model, Clustering-aid decomposition and merging (c-DM) algorithm | 1) Modeled the fleet dispatching problem for mobile electric-vehicle-to-electric-vehicle (mE2) charging on the move technology mathematically on an encounter network with trip synchronization and unique illegal subtour elimination.<br>2) Addressed the scalability difficulties of mE2 charging problem with a new c-DM heuristic algorithm.<br>3) Explored the potential benefits and feasibility of mE2 charging technology in practical scenarios on Chicago and Florida networks. |

In summary, the major contributions of our study are shown as follows:
1) We mathematically formulate a mixed integer linear programming model for the platoon-based vehicle-to-vehicle charging (PV2VC) problem, along with two problem variants for the electric vehicle routing problem (EVRP) and electric vehicle platooning problem (EVPP).
2) A customized genetic algorithm (GA) is proposed to solve large-scale PV2VC problem in practical scenarios, including the EVRP and EVPP variants.
3) We conduct numerical experiments to evaluate the computational performance of GA and explore the potential benefits of PV2VC technology under various operational conditions.

## 3. Mathematical model

Given a fleet of ER and ES vehicles with their initial locations, battery levels and capacities, and sequences of visiting locations for ERs, the objective is to find the optimal routes and charging schedules to minimize the total cost of ERs. The total cost is measured by the weighted sum of energy consumption and travel time for ERs, where the travel time might include the charging time at CS, platoon formation delay, and wait time for ES for the PV2VC service. The optimal solution can include all or any combination of the three operating scenarios introduced in Section 2.1. One ER is allowed to travel in platoon along part of the trips with different vehicles, detour to CS to get partially charged, and receive the PV2VC service en-route by multiple ESs as well.

## 3.1 Proposed formulation

The PV2VC problem is defined on a directed graph $G(N, A)$ over an operational time horizon $[0, T]$. $N$ is a set of nodes, consisting of vehicle initial locations, visiting tasks of ERs, platoon join/split locations, and charging stations. $N^{CS}$ is the set of charging station locations and $N'$ is the set of all other nodes ($N = \{N^{CS} \cup N'\}$). $\lambda_i$ represents the charging rate if $i \in N^{CS}$, or equals to 0 if $i \in N'$. We use $\sigma_i$ to represent the minimum required battery level to the nearest CS from node $i$ for ES. $A$ is the set of directed links, where $A_i^+$ and $A_i^-$ represent the set of inbound and outbound arcs at node $i$, respectively. For each arc $ij \in A$, we use $d_{ij}$ to denote the travel distance, $\tau_{ij}$ for the travel time, and $c_{ij}$ for the energy consumption rate regarding $d_{ij}$. The notation is shown in **Table 3**.

Table 3. Model notations

| Notations | Definitions |
|---|---|
| *Parameters* | |
| $d_{ij}$ | travel distance for arc $ij \in A$ |
| $\tau_{ij}$ | travel time for arc $ij \in A$ |
| $c_{ij}$ | energy consumption rate for arc $ij \in A$ |
| $\lambda_i$ | charging rate at charging station (CS) if $i \in N^{CS}$, or 0 if $i \in N'$ |
| $\sigma_i$ | minimum required battery level to the nearest CS at node $i \in N$ |
| $R$ | set of electricity requests (ER) |
| $\Delta_r$ | list of visiting nodes of $r \in R$, starting from origin ($\delta_r^O$) to destination ($\delta_r^D$) through a sequence of intermediate nodes |
| $\varphi_r$ | minimum required battery level of request $r \in R$ |
| $S$ | set of electricity suppliers (ES) |
| $\omega_s$ | V2V power transfer rate of supplier $s \in S$ |
| $K$ | $K = \{R \cup S\}$ |
| $\delta_k^O$ | origin of vehicle $k \in K$ |
| $\delta_k^D$ | destination (dummy depot for ES) of vehicle $k \in K$ |
| $t_k$ | request submission time (ER) or the earliest ready time (ES) at origin of $k \in K$ |
| $b_k$ | battery capacity of $k \in K$ |
| $e_k$ | initial state of charge $k \in K$ |
| $\eta_1$ | platoon saving rate |
| $\eta_2$ | V2V power transfer efficiency |
| $M$ | large constant number |
| *Decision variables* | |
| $X_{ijk}$ | binary: 1 if vehicle $k \in K$ traverses arc $ij$, and 0 otherwise |
| $T_{ik}$ | continuous: time at which vehicle $k \in K$ arrives at node $i$ |
| $U_{ik}$ | continuous: charging time at CS if $i \in N^{CS}$ or wait time if $i \in N'$ |
| $E_{ik}$ | continuous: battery level of vehicle $k \in K$ at node $i$ |
| $F_{ijk}$ | binary: 1 if vehicle $k \in K$ traverses arc $ij$ in platoon |
| $P_{ijrk}$ | binary: 1 if request $r \in R$ and $k \in K$ traverse arc $ij$ in platoon |
| $Q_{ijrs}$ | continuous: percentage of time that request $r \in R$ get charged from supplier $s \in S$ while traverse arc $ij$ together |
| *Auxiliary variables* | |
| $Z_{ijk}$ | dummy variable for the left-hand-side of nonlinear constraints |
| $W_{ijk}$ | dummy variable for the right-hand-side of nonlinear constraints |

Let $R$ be the set of electricity requests, $S$ be the set of electricity suppliers, and $K = \{R \cup S\}$ be the set of all ERs and ESs. We use $\Delta_r$ to represent the sequence of visiting nodes of $r \in R$, where it starts from an initial location to the destination through several intermediate nodes. $\varphi_r$ is the minimum required battery level of $r \in R$. $\omega_s$ denotes the vehicle-to-vehicle power transfer rate of $s \in S$.

For any vehicle $k \in K$, the origin and destination (idle ESs are assigned to "dummy depots" representing designated idling locations) are $\delta_k^O$ and $\delta_k^D$. The dummy depot is employed to allow the ES to end its service at any location on the network as a place for relocating idle vehicles for future requests. An ES ending up at a dummy depot indicates that it is ready for new service in future time intervals under dynamic settings. By default, we assume that every node on the network is connected to the dummy depot directly without any cost. If an ES is not assigned to any ER, it would move directly to the dummy depot without any cost. For more practical scenarios that integrate vehicle relocation problem (Pantelidis et al., 2022), multiple dummy depots can be assigned on the network to represent different EV charging facilities, transit stations or warehouses with different relocation costs. For simplicity, the dummy depot is not shown on the graph. We use $t_k$ for the request submission time of ER and the earliest ready time at initial location of ES. $b_k$ represents the battery capacity of $k \in K$.

The proposed MILP model is shown in constraints (1) – (7). Integral and non-negativity constraints are defined in constraints (8) to (16).

$$\min \alpha \sum_{r \in R} \sum_{(i,j) \in A} c_{ij} d_{ij} (X_{ijr} - \eta_1 F_{ijr}) + \beta \sum_{r \in R} (T_{\delta_r^D r} - t_r) \quad (1)$$

Subject to

$$\sum_{(i,j) \in A_i^-} X_{ijr} = 1, \quad \forall r \in R, \forall i \in \Delta_r \setminus \{\delta_r^D\} \quad (2)$$

$$\sum_{(j,i) \in A_i^+} X_{jir} = 1, \quad \forall r \in R, i = \delta_r^D \quad (3)$$

$$\sum_{(i,j) \in A_i^-} X_{ijr} - \sum_{(j,i) \in A_i^+} X_{jir} = 0, \quad \forall r \in R, \forall i \in N \setminus \{\delta_r^O, \delta_r^D\} \quad (4)$$

$$\sum_{(i,j) \in A_i^-} X_{ijs} = 1, \quad \forall s \in S, i = \delta_s^O \quad (5)$$

$$\sum_{(j,i) \in A_i^+} X_{jis} = 1, \quad \forall s \in S, i = \delta_s^D \quad (6)$$

$$\sum_{(i,j) \in A_i^-} X_{ijs} - \sum_{(j,i) \in A_i^+} X_{jis} = 0, \quad \forall s \in S, \forall i \in N \setminus \{\delta_s^O, \delta_s^D\} \quad (7)$$

$$X_{ijk} \in \{0,1\}, \quad \forall k \in K, \forall (i,j) \in A \quad (8)$$
$$T_{ik} \geq 0, \quad \forall k \in K, \forall i \in N \quad (9)$$
$$U_{ik} \geq 0, \quad \forall k \in K, \forall i \in N \quad (10)$$
$$E_{ik} \geq 0, \quad \forall k \in K, \forall i \in N \quad (11)$$
$$F_{ijk} \in \{0,1\}, \quad \forall k \in K, \forall (i,j) \in A \quad (12)$$
$$P_{ijrk} \in \{0,1\}, \quad \forall r, k \in K, r \neq k, \forall (i,j) \in A \quad (13)$$
$$Q_{ijrs} \in \{0,1\}, \quad \forall r \in R, s \in S, \forall (i,j) \in A \quad (14)$$
$$Z_{ijk} \geq 0, \quad \forall k \in K, \forall (i,j) \in A \quad (15)$$
$$W_{ijk} \geq 0, \quad \forall k \in K, \forall (i,j) \in A \quad (16)$$

Objective function (1) minimizes the total of energy consumption and travel time for ER, weighted by parameters $\alpha$ and $\beta$ in general form. The weights $\alpha$ and $\beta$ are set by the operator. As shown in the objective function, the vehicle platoon savings are deducted by a constant rate of $\eta_1$ from the energy consumption. Although the average travel cost for vehicles in a platoon decreases nonlinearly with the platoon length (Tsugawa et al., 2016; Song et al., 2021) and many studies in the literature adopt a linear relationship for the platoon cost saving rate (Larsson et al., 2015; Boysen et al., 2018; Fu and Chow, 2023), we assume that vehicles traveling in platoons only receive a constant saving rate in energy consumption for simplicity. The second term of the objective function measures the travel time of the ER through the entire trip, which is calculated by the difference between the arrival time at destination and the time when the request is submitted. Thus, the travel time might include the charging time at CS, platoon formation delay, and wait time for the ES for the PV2VC service.

Constraints (2) ensure that each ER leaves its initial location and every intermediate node in $\Delta_r$. Constraints (3) ensure that the ER ends its trip at the destination. Constraints (4) maintain the vehicle flow conservation at any node except the origin and destination locations of ER. Constraints (2) and (4) together guarantee that every intermediate node in $\Delta_r$ is visited. Constraints (5) and (6) ensure that each ES leaves its initial location and ends the trip at the dummy depot. Constraints (7) keep the vehicle flow conservation at all nodes except the origin and dummy depot.

## 3.2   Vehicle arrival time and precedence of ER tasks

To measure the travel time of ER, we adopt constraints (17) to track the vehicle arrival time along the path. They also work as subtour elimination constraints. The time windows at the initial location are enforced by constraints (18), where the arrival time at the origin has to be greater than the request submission time for the ER and the earliest ready time for the ES. Constraints (19) ensure the precedence of visiting tasks for $r \in R$.

$$T_{ik} + \tau_{ij} X_{ijk} + U_{ik} \leq T_{jk} + M(1 - X_{ijk}), \quad \forall (i,j) \in A, \forall k \in K \tag{17}$$

$$T_{ik} \geq t_k, \quad \forall k \in K, i = \delta_k^O \tag{18}$$

$$T_{i_n r} \leq T_{i_{n+1} r}, \\ \forall r \in R, i_n, i_{n+1} \in \Delta_r, \Delta_r = \{\cdots, i_n, i_{n+1}, \cdots\} \tag{19}$$

## 3.3   Platoon identification

We assume that an ES can only provide the PV2VC service to one ER over an arc $ij$ at the same time like in the literature. Thus, we can identify the matching and quantify the electricity transfer between ES and ER with this one-to-one relationship assumption. However, an ER is allowed to travel with multiple ERs at the same time. Moreover, if one ER is matched and served with an ES, we then consider that the associated ES is also traveling in the same platoon. Therefore, we employ a binary decision variable $P_{ijrk}$, which equals 1 if an ER $r \in R$ is traveling in a platoon with another vehicle $k \in K$ ($r \neq k$) over arc $ij$, and 0 otherwise. Constraints (20) and (21) ensure that the pairwise platoon vehicles have the same departure time at node $i$. Similarly, constraints (22) and (23) ensure the same arrival time at node $j$. Constraints (23) make sure that $P_{ijrk}$ equals to 1 if and only if both vehicles are traveling on arc $ij$.

$$(T_{ir} + U_{ir}) - (T_{ik} + U_{ik}) \leq M(1 - P_{ijrk}), \quad \forall (i,j) \in A, \forall r \in R, \forall k \in K, r \neq k \tag{20}$$

$$(T_{ik} + U_{ik}) - (T_{ir} + U_{ir}) \leq M(1 - P_{ijrk}), \quad \forall (i,j) \in A, \forall r \in R, \forall k \in K, r \neq k \tag{21}$$

$$T_{jr} - T_{jk} \leq M(1 - P_{ijrk}), \quad \forall (i,j) \in A, \forall r \in R, \forall k \in K, r \neq k \tag{22}$$

$$T_{jk} - T_{jr} \leq M(1 - P_{ijrk}), \quad \forall (i,j) \in A, \forall r \in R, \forall k \in K, r \neq k \tag{23}$$

$$2 P_{ijrk} \leq X_{ijr} + X_{ijk}, \quad \forall (i,j) \in A, \forall r \in R, \forall k \in K, r \neq k \tag{24}$$

Constraints (25) and (26) indicate whether the ER $r \in R$ or ES $s \in S$ travels in a platoon on arc $ij$, respectively. The binary decision variable $F_{ijk}$ is set to 1 to reduce the energy consumption in objective function (1) only if it travels in platoon with other vehicles (captured by the corresponding $P_{ijrk}$). Constraints (27) and (28) ensure the one-on-one relationship between ER and ES. In other words, each ER can only accept electricity from one ES and each ES can only provide electricity to one ER at a time on the same arc.

$$F_{ijr} \leq \sum_{k \in K, r \neq k} P_{ijrk}, \quad \forall (i,j) \in A, \forall r \in R \tag{25}$$

$$F_{ijs} \leq \sum_{r \in R} P_{ijrs}, \quad \forall (i,j) \in A, \forall s \in S \tag{26}$$

$$\sum_{s \in S} P_{ijrs} \leq 1, \quad \forall\,(i,j) \in A, \forall r \in R \tag{27}$$

$$\sum_{r \in R} P_{ijrs} \leq 1, \quad \forall\,(i,j) \in A, \forall s \in S \tag{28}$$

### 3.4 Energy conservation

To explicitly quantify the exact amount of energy consumed, transferred, and received during the trips, we propose constraints (29) and (30) for the energy conservation of ER and ES, respectively. If the ER or ES do not travel on the arc $ij$ ($X_{ijr} = 0$, $X_{ijs} = 0$), constraints (29) and (30) can be ignored. If $X_{ijr}$ equals to 1 in (29), the left side represents for ER $r \in R$: the energy upon arrival at node $i$, plus the electricity charged at node $i$ if $i \in N^{CS}$ and $U_{ir} > 0$, plus the energy received from the PV2VC service over arc $ij$, minus the energy consumption across the arc $ij$ (in a platoon or by itself). The integration of all left side terms is equivalent to the energy upon arrival at node $j$, which is the right side of constraints (29).

Similarly, the left side terms of constraints (30) represent the energy conservation for ES $s \in S$ between node $i$ and node $j$: the arrival energy at node $i$, plus the possible electricity charged at node $i$, minus the energy transferred out for the PV2VC service and the energy consumption across the arc $ij$. The decision variable $Q_{ijrs}$ is between 0 to 1 and denotes the percentage of time on $\tau_{ij}$ that request $r \in R$ charges from supplier $s \in S$ when they traverse the arc $ij$ together.

$$X_{ijr}\left(E_{ir} + \lambda_i U_{ir} + \eta_2 \tau_{ij} \sum_{s \in S} \omega_s Q_{ijrs} - c_{ij} d_{ij}(X_{ijr} - \eta_1 F_{ijr})\right) = X_{ijr}(E_{jr}), \tag{29}$$
$$\forall (i,j) \in A, \forall r \in R$$

$$X_{ijs}\left(E_{is} + \lambda_i U_{is} - \tau_{ij} \sum_{r \in R} \omega_s Q_{ijrs} - c_{ij} d_{ij}(X_{ijs} - \eta_1 F_{ijs})\right) = X_{ijs}(E_{js}), \tag{30}$$
$$\forall (i,j) \in A, \forall s \in S$$

Constraints (29) and (30) are expressed in nonlinear forms but they can be replaced with equivalent inequalities as linear forms (see Fu and Chow, 2022). The constraints can be replaced by a form shown in Eqs. (31) to (33), where $Z$ is a continuous decision variable, and A is another continuous variable bounded by $[0, M)$ with $M$ as a large constant, and $x$ is a binary variable.

$$Z \leq Mx \tag{31}$$
$$Z \leq A \tag{32}$$
$$Z \geq A - M(1 - x) \tag{33}$$

By applying the inequalities (31) to (33), constraints (29) are replaced by Eq. (34), where the terms are replaced by constraints (35) to (37) and constraints (38) to (40), respectively.

$$Z_{ijr} = W_{ijr}, \quad \forall (i,j) \in A, \forall r \in R \tag{34}$$
$$Z_{ijr} \leq MX_{ijr}, \quad \forall (i,j) \in A, \forall r \in R \tag{35}$$
$$Z_{ijr} \leq E_{ir} + \lambda_i U_{ir} + \eta_2 \tau_{ij} \sum_{s \in S} \omega_s Q_{ijrs} - c_{ij} d_{ij}(X_{ijr} - \eta_1 F_{ijr}) \tag{36}$$
$$\forall (i,j) \in A, \forall r \in R$$
$$Z_{ijr} \geq E_{ir} + \lambda_i U_{ir} + \eta_2 \tau_{ij} \sum_{s \in S} \omega_s Q_{ijrs} - c_{ij} d_{ij}(X_{ijr} - \eta_1 F_{ijr}) - M(1 - X_{ijr}) \tag{37}$$
$$\forall (i,j) \in A, \forall r \in R$$
$$W_{ijr} \leq MX_{ijr}, \quad \forall (i,j) \in A, \forall r \in R \tag{38}$$
$$W_{ijr} \leq E_{jr}, \quad \forall (i,j) \in A, \forall r \in R \tag{39}$$
$$W_{ijr} \geq E_{jr} - M(1 - X_{ijr}), \quad \forall (i,j) \in A, \forall r \in R \tag{40}$$

Similarly, constraints (30) are replaced by Eqs. (41) to (47).

$$Z_{ijs} = W_{ijs}, \quad \forall (i,j) \in A, \forall s \in S \tag{41}$$

$$Z_{ijs} \leq M X_{ijs}, \quad \forall (i,j) \in A, \forall s \in S \tag{42}$$

$$Z_{ijs} \leq E_{is} + \lambda_i U_{is} - \tau_{ij} \sum_{r \in R} \omega_s Q_{ijrs} - c_{ij} d_{ij} (X_{ijs} - \eta_1 F_{ijs})$$
$$\forall (i,j) \in A, \forall s \in S \tag{43}$$

$$Z_{ijs} \geq E_{is} + \lambda_i U_{is} - \tau_{ij} \sum_{r \in R} \omega_s Q_{ijrs} - c_{ij} d_{ij} (X_{ijs} - \eta_1 F_{ijs}) - M(1 - X_{ijs})$$
$$\forall (i,j) \in A, \forall s \in S \tag{44}$$

$$W_{ijs} \leq M X_{ijs}, \quad \forall (i,j) \in A, \forall s \in S \tag{45}$$

$$W_{ijs} \leq E_{js}, \quad \forall (i,j) \in A, \forall s \in S \tag{46}$$

$$W_{ijs} \geq E_{js} - M(1 - X_{ijs}), \quad \forall (i,j) \in A, \forall s \in S \tag{47}$$

Constraints (48) ensure the limit of $Q_{ijrs}$. The minimum and maximum battery capacity limits for ER and ES are guaranteed by constraints (49) to (50) and constraints (51) to (52), respectively. The battery levels for ER and ES are set to $e_k$ at initial locations by constraints (53).

$$0 \leq Q_{ijrs} \leq P_{ijrs}, \quad \forall (i,j) \in A, \forall r \in R, \forall s \in S \tag{48}$$

$$E_{ir} \geq \varphi_r, \quad \forall i \in N, \forall r \in R \tag{49}$$

$$E_{ir} + \lambda_i U_{ir} \leq b_r, \quad \forall i \in N, \forall r \in R \tag{50}$$

$$E_{is} \geq \sigma_i, \quad \forall i \in N, \forall s \in S \tag{51}$$

$$E_{is} + \lambda_i U_{is} \leq b_s, \quad \forall i \in N, \forall s \in S \tag{52}$$

$$E_{ik} = e_k, \quad \forall k \in K, i = \delta_k^O \tag{53}$$

### 3.5 Problem variants and model summary

In summary, the entire MILP model for the PV2VC problem consists of Eqs. (1) – (28), (34) – (53). However, to evaluate the computation performance and potential benefits of PV2VC, we need to compare against two other problem variants, EVRP and EVPP. Thus, we provide minor modifications in this section to simply the MILP model to address them.

By replacing the objective function (1) with (1.b) and the constraints (29) with a simplified version as (29.b), the MILP model for the EVRP now consists of Eqs. (1.b), (2) – (4), (8) – (11), (15) – (19), (34) – (35), (36.b) – (37.b), (38) – (40), (49) – (50), and (53).

$$\min \alpha \sum_{r \in R} \sum_{(i,j) \in A} c_{ij} d_{ij} X_{ijr} + \beta \sum_{r \in R} (T_{\delta_r^D r} - t_r) \tag{1.b}$$

$$X_{ijr}(E_{ir} + \lambda_i U_{ir} - c_{ij} d_{ij} X_{ijr}) = X_{ijr}(E_{jr}), \quad \forall (i,j) \in A, \forall r \in R \tag{29.b}$$

$$Z_{ijr} \leq E_{ir} + \lambda_i U_{ir} - c_{ij} d_{ij} X_{ijr}, \quad \forall (i,j) \in A, \forall r \in R \tag{36.b}$$

$$Z_{ijr} \geq E_{ir} + \lambda_i U_{ir} - c_{ij} d_{ij} X_{ijr} - M(1 - X_{ijr}), \quad \forall (i,j) \in A, \forall r \in R \tag{37.b}$$

Similarly, we replace the constraints (29) with (29.c) and the complete MILP model for the EVPP now includes Eqs. (1) – (4), (8) – (13), (15) – (25), (34) – (35), (36.c) – (37.c), (38) – (40), (49) – (50), and (53).

$$X_{ijr}\left(E_{ir} + \lambda_i U_{ir} - c_{ij} d_{ij}(X_{ijr} - \eta_1 F_{ijr})\right) = X_{ijr}(E_{jr}), \quad \forall (i,j) \in A, \forall r \in R \tag{29.c}$$

$$Z_{ijr} \leq E_{ir} + \lambda_i U_{ir} - c_{ij} d_{ij}(X_{ijr} - \eta_1 F_{ijr}), \quad \forall (i,j) \in A, \forall r \in R \tag{36.c}$$

$$Z_{ijr} \geq E_{ir} + \lambda_i U_{ir} - c_{ij} d_{ij}(X_{ijr} - \eta_1 F_{ijr}) - M(1 - X_{ijr}), \quad \forall (i,j) \in A, \forall r \in R \tag{37.c}$$

As the PV2VC problem can be simplified to a vehicle routing problem, the problem is NP-hard and requires efficient heuristic algorithms to solve instances of practical size.

## 4. Genetic algorithm

To solve the PV2VC for practical scenarios and compare against the EVRP and EVPP, we propose a class of genetic algorithms (GAs) for all three variants. A typical structure of a GA includes the genetic representation of the solution, generation of initial population, fitness function, selection, and a series of genetic operators for new generations such as crossover and mutation. In this section, we extend and develop two types of genetic operators based on the features of crossover and mutation respectively: 1) modify the vehicle path to locate the charging stations and platoon join/split locations, 2) modify the charging time at charging stations and the percentage of charging time for the PV2VC service.

The proposed GA is presented in three parts, addressing the three variants, respectively: the EVRP, EVPP, and PV2VC problem.

### *4.1 GA for the electric vehicle routing problem*

The overall framework of the proposed GA for the EVRP is shown in **Figure 3**. The GA would terminate if it either reaches the maximum number of generations or the limit of iterations without improvement.

#### *4.1.1 Solution representation*

In the electric vehicle routing problem, each individual vehicle can be considered as an independent sub-problem. We use two sets of chromosomes to represent the solution to the EVRP. For an electricity request $r \in R$, the first layer of a solution contains a sequence of ordered visiting nodes for the EV, denoted as a set $L_r^1$. The second layer of a solution represents the dwell time at each visiting node or charging time if the associated node is a charging station ($N^{CS}$), denoted as a set $L_r^2$. For any $l_r^1 \in L_r^1$ and $l_r^2 \in L_r^2$, $l_r^1$ and $l_r^2$ always have the same length. $L_r^2$ always equals to 0 at non-CS nodes ($N'$) for EVRP. For example, an EV may have a vehicle path ($l_r^1$) as: $0 \to 3 \to 1 \to 4 \to 2$, and a dwell time result ($l_r^2$) as: $0 \to 0 \to 30 \to 0 \to 0$, which indicates that the EV only stays at node 1 for 30 mins for charging (assume node 1 is a CS).

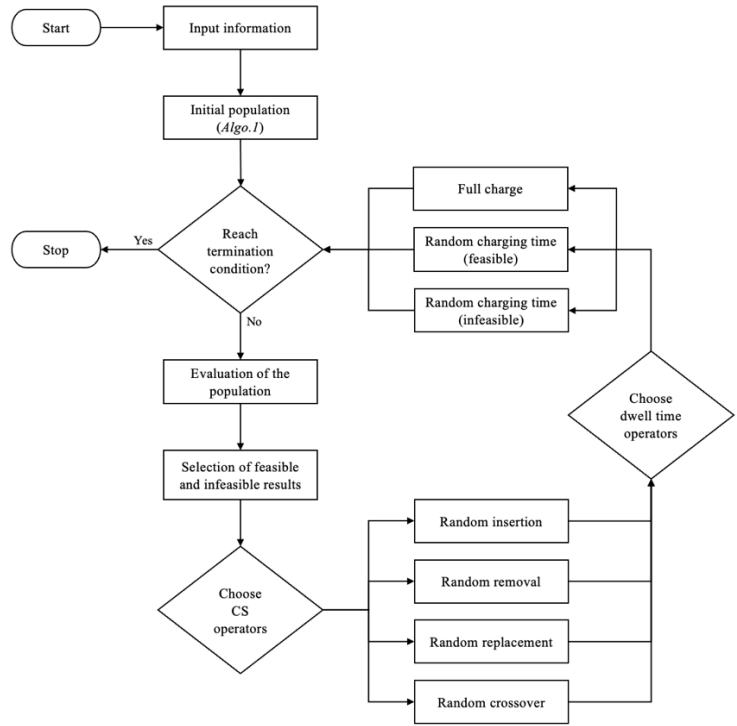

Figure 3. Flowchart of proposed GA for EVRP

### 4.1.2 Initial population, evaluation and selection

Before generating the initial population, we first check the feasibility of a given EV route ($\Delta_r$). If the EV can neither finish its assigned tasks nor reach the closest charging station from its initial location ($\delta_r^O$), this EV does not have a feasible solution. Moreover, if the EV can finish all assigned tasks without any en-route charging, we output the given EV path directly. For all other cases, we develop a simple construction heuristic as shown in **Algorithm 1** (*Algo.1*) to generate the set of initial solutions.

**Algorithm 1.** Construction heuristic for initial population

| | |
|---|---|
| Input: | Graph $G(N,A)$ with $c_{ij}$ and $\lambda_i$, request $r$ with $\Delta_r$, $\varphi_r$, $t_r$, $b_r$, $e_r$. |
| Initialization: | Shortest-path distance matrix $D_{ij}$ for $\forall i,j \in N$. Dwell time $l_r^2$ all set to 0. Full battery charging time from 0 to $b_r$ as $t_r^{max}$. |
| 1. | **If** $r$ can neither finish $\Delta_r$ nor reach any $n \in N^{CS}$ from $\delta_r^O$, **Return** infeasible. |
| 2. | **If** $r$ can finish $\Delta_r$ without charging, **Return** feasible, $\Delta_r$ and $l_r^2$. |
| 3. | **For** any adjacent $i_n, i_{n+1} \in \Delta_r$: |
| 4. |    Identify the $n^* \in N^{CS}$ with minimum detour connecting $i_n$ and $i_{n+1}$. |
| 5. |    **If** $r$ can reach $n^*$ from $\delta_r^O$ without charging and $E_{n^*r} \geq \varphi_r$: |
| 6. |      Insert $n^*$ in-between $i_n$ and $i_{n+1}$ and modify $\Delta_r$ as new $l_r^1$. |
| 7. |      Assign four new $l_r^2$, each with 25%, 50%, 75% and 100% of $t_r^{max}$ at $n^*$. |
| 8. |      Record four new sets of solutions in $L_r^1$ and $L_r^2$. |
| 9. |    **Else:** |
| 10. |      **Return** feasible, $L_r^1$ and $L_r^2$. |

To evaluate the solution quality, we maintain the same objective as the MILP model and measure the total energy consumption and travel time from $L_r^1$ and $L_r^2$ as the fitness value. For feasible solutions, the lower the fitness value the better the solution. As for infeasible solutions, we instead measure the energy shortage as a penalty but still record the result as they might improve and become feasible in subsequent generations. Among all solutions, we select a constant number of feasible results with the lowest fitness value and infeasible results with the lowest penalty value for next generations.

### 4.1.3 Genetic operators

For creating new solutions through generations, we have two sets of genetic operators. The first type of operator modifies the locations of charging station in the visiting sequence of any $l_r^1 \in L_r^1$.

- *Random insertion*: Select a random set of adjacent nodes $i_n, i_{n+1} \in l_r^1$ and insert a CS $n^* \in N^{CS}$ with minimum detours. Update $l_r^1$ and associated $l_r^2$.
- *Random removal*: Select a random CS in $l_r^1$ for removal. Update $l_r^1$ and associated $l_r^2$. Skip if there is no existing CS in $l_r^1$.
- *Random replacement*: Select a random CS in $l_r^1$ and replace it with the closest CS nearby. Update $l_r^1$ and associated $l_r^2$. Skip if there is no existing CS in $l_r^1$.
- *Random crossover*: Select a random CS in $l_r^1$ and switch the location with one of the adjacent visiting nodes in $l_r^1$. Update $l_r^1$ and associated $l_r^2$. Skip if there is no existing CS in $l_r^1$.

The second type of genetic operators changes the dwell time (in mins) at charging stations in any $l_r^2 \in L_r^2$. We skip if there is no existing CS in $l_r^2$. To make trade-offs between the solution speed and accuracy, we limit the generated dwell time to a specific decimal place before running the GA.

- *Full charge*: Select a random CS and set the associated dwell time to $t_r^{max}$.
- *Random charging time (feasible)*: If current $l_r^1$ and $l_r^2$ is feasible, select a random CS and randomly generate a new dwell time between 0 to current dwell time at the CS.

- *Random charging time (infeasible)*: If current $l_r^1$ and $l_r^2$ is infeasible, select a random CS and randomly generate a new dwell time between current dwell time at the CS to $t_r^{max}$.

## 4.2 Electric vehicle platooning problem

For solving the EVPP, we adopt the similar solution framework as Fu and Chow (2023), which also focuses on the vehicle platooning problem. There are three major steps in this section. First, the algorithm searches for potential platoons for any two vehicles based on the results from EVRP. Once more than two feasible electric vehicle platoons are found, the algorithm attempts to merge them together if they share a common platoon path. Lastly, remaining individual vehicles (RIVs) that are not assigned to any platoon yet are considered for insertion into existing platoons. In each major step, the algorithm selects the EV platoon assignments and schedules that save the most costs from the EVRP results. The overall GA framework for EVPP is shown in **Figure 4**.

### 4.2.1 Solution representation

In addition to the two layers introduced in Section 4.1.1, we have another two layers in solution representation to solve the EVPP. For an electricity request $r \in R$, the third layer ($L_r^3$) records all the vehicles that travel in platoon with $r$, including $r$ itself. The fourth layer ($L_r^4$) represents the arrival time along the vehicle path of an ER, which is used for the spatial-temporal synchronization of platoon join/split operations and the measurement of total travel time. For example, an EV may have a platoon result ($l_r^3$) as: None $\rightarrow [r, x, y] \rightarrow [r, x, y] \rightarrow [r, x, y] \rightarrow$ None, and an arrival time result ($l_r^4$) as: $0 \rightarrow 15 \rightarrow 20 \rightarrow 50 \rightarrow 70$. This indicates that the EV travels in platoon with two other vehicles, $x$ and $y$, from the second to the fourth visiting nodes, and the overall travel time is equal to 70 min assuming that $r$ submits the request at time 0.

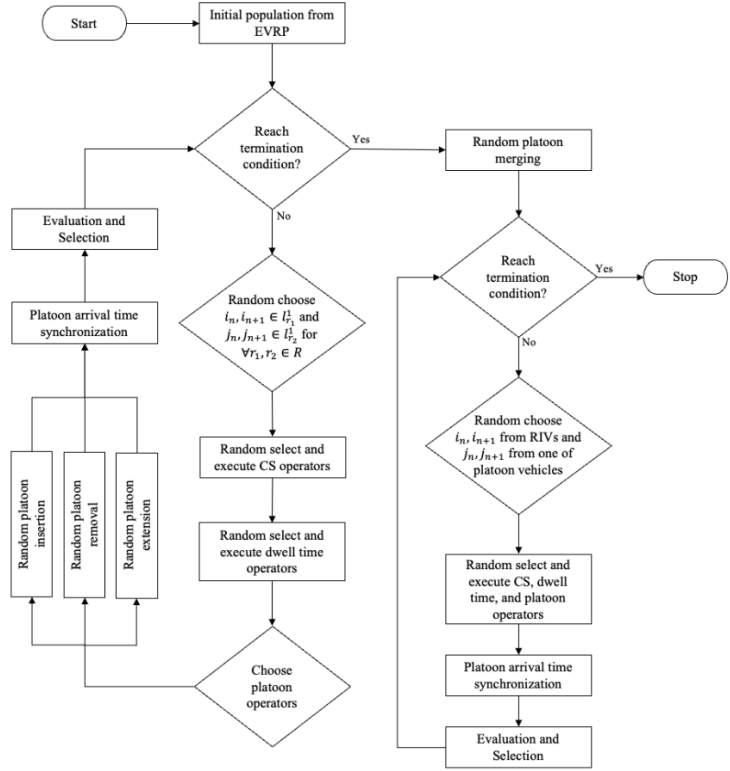

Figure 4. Flowchart of proposed GA for EVPP

### 4.2.2 Initial population, evaluation and selection

The results from the EVRP are recorded and used as the initial population for the EVPP in the proposed genetic algorithm. For the fitness function, total energy consumption and travel time for each individual vehicle is tracked. For any ER $r \in R$, the algorithm calculates the total energy consumption from the assignments of $L_r^1$ and $L_r^3$, which includes the energy savings from platooning. As for the total travel time, it is calculated from the arrival time at the destination on the fourth layer ($L_r^4$). If the assignments of $L_r^1$, $L_r^2$, $L_r^3$ and $L_r^4$ are not feasible, the energy shortage is applied as a penalty. Among all feasible platoon results, the algorithm selects the platoon with the most savings relative to the initial EVRP results.

### 4.2.3 Genetic operators

Except for the basic charging station and dwell time operators, a set of platoon-specific operators are designed for the EVPP. Moreover, the arrival times ($L_r^4$) for all vehicles in the same platoon are synchronized based on the assignments of $L_r^1$, $L_r^2$, and $L_r^3$.

- *Random platoon insertion*: Given two ERs $r_1, r_2 \in R$, random select candidate platoon insertion segments $i_n, i_{n+1} \in l_{r_1}^1$ and $j_n, j_{n+1} \in l_{r_2}^1$. This operator is shown in **Algorithm 2** (*Algo.2*).
- *Random platoon removal*: Given a platoon of vehicles, random remove an existing platoon. Update associated $L_r^1$, $L_r^2$, and $L_r^3$. Skip if there are no existing platoons.
- *Random platoon extension*: Given a platoon of vehicles, randomly select an existing platoon and choose to extend either the join or split side. This operator is similar as *Algo.2* but either $i_n = j_n$ or $i_{n+1} = j_{n+1}$ in this case. Update associated $L_r^1$, $L_r^2$, and $L_r^3$. Skip if there are no existing platoons.
- *Random platoon merging*: Given two platoons, randomly merge them on the common platoon segments. Update associated $L_r^1$, $L_r^2$, and $L_r^3$. Skip if there are no common platoon segments between them.

**Algorithm 2.** Random platoon insertion

| | |
|---|---|
| Input: | Two requests $r_1, r_2 \in R$ with associated $l_{r_1}^1, l_{r_1}^2, l_{r_1}^3$ and $l_{r_2}^1, l_{r_2}^2, l_{r_2}^3$. |
| | Candidate platoon insertion segments $i_n, i_{n+1} \in l_{r_1}^1$ and $j_n, j_{n+1} \in l_{r_2}^1$. |
| 1. | Identify the nodes along the shortest path of $(i_n, i_{n+1})$ as $p_{r_1}$ and $(j_n, j_{n+1})$ as $p_{r_2}$. |
| 2. | Random select $i'_n, i'_{n+x} \in p_{r_1}$ and $j'_n, j'_{n+y} \in p_{r_2}$. |
| 3. | Identify the nodes with min total distance and difference value to $(i'_n, j'_n)$ as platoon join node $i^*$ and $(i'_{n+x}, j'_{n+y})$ as platoon split node $j^*$. |
| 4. | Insert $(i^*, j^*)$ in-between $(i_n, i_{n+1})$ in $l_{r_1}^1$ and $(j_n, j_{n+1})$ in $l_{r_2}^1$ as the platoon segment. |
| 5. | **Return** modified results of $l_{r_1}^1, l_{r_1}^2, l_{r_1}^3$ and $l_{r_2}^1, l_{r_2}^2, l_{r_2}^3$. |

Note: $x, y$ are randomly selected positive constants.
$n + x$ and $n + y$ are less than the length of $p_{r_1}$ and $p_{r_2}$, respectively.

### 4.3 Platoon-based vehicle-to-vehicle charging (PV2VC) problem

In this section, electricity suppliers are introduced, and they are given at initial locations $\delta_s^O$ (mostly at CS) with an initial state of charge $e_s$. These ESs can be dispatched, travel in platoon with ERs towards their destinations, and provide charging services while moving.

Similar to the EVPP, there are four major steps for solving the PV2VC problem. First, a one-to-one ER-ES assignment is applied to the EVRP solutions to reduce the energy consumption and travel time of ERs. Then, if there are more than two feasible ER-ES assignments and they share a common platoon path, the algorithm tries to merge them into a larger platoon. Third, RIVs that are not matched with any ES yet are considered for insertion into existing ER-ES platoons. The new insertion of an ER may or may not get the PV2VC service from the ES and could just travel in platoon with the ERs in the platoon. In the last step, if there are still RIVs left, they are simply tested for platooning only as EVPP. The GA framework customized for the PV2VC problem is shown in **Figure 5**.

### 4.3.1 Solution representation

Since the ES is introduced in this section, minor modifications are made to the four layers of solution representation. For the ER, the third layer $L_r^3$ might also include ESs in their solutions. In terms of the ES, they have $L_s^1$ for the vehicle path, $L_s^2$ for the dwell time at each location, and $L_s^4$ for the arrival time along the route. But the third layer $L_s^3$ only includes the ER which receives the PV2VC service to align with the assumption and constraint that one ES can only charge one ER at a time. Based on the assignment of $L_s^3$, one more solution layer ($L_s^5$) is added only for the ES, which indicates the percentage of PV2VC service time when traveling with the associated ER over the segment.

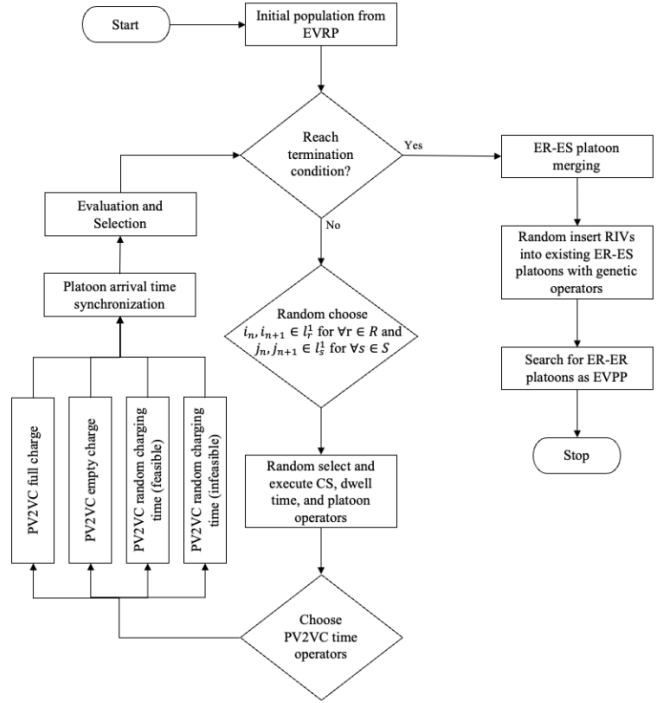

Figure 5. Flowchart of proposed GA for PV2VC problem

One complete solution representation example for the PV2VC problem is shown in 
**(a)** (b)
Figure 6. In this example, the ER ($r$) has a vehicle path $L_r^1$: 2 → 3 → 1 → 4, and ES ($s$) has a vehicle path $L_s^1$: 0 → 3 → 1, where the ER receives the PV2VC service from the ES on the segment of 3 → 1. Note that the ER arrives at the platoon join location (node 3) at time 30 ($L_r^4$), which is later than the ES that arrives at time 25 ($L_s^4$). Thus, the ES waits at node 3 for an additional 5 mins as noted in $L_s^2$. According to the fifth layer of ES ($L_s^5$), it charges the ER for half of the time while traversing the segment of 3 → 1 in platoon together with the ER.

| $L_r^1$ : | 2 | 3 | 1 | 4 |
|---|---|---|---|---|
| $L_r^2$ : | 10 | 0 | 25 | 0 |
| $L_r^3$ : | None | s | s | None |
| $L_r^4$ : | 0 | 30 | 40 | 75 |

| $L_s^1$ : | 0 | 3 | 1 |
|---|---|---|---|
| $L_s^2$ : | 0 | 5 | 0 |
| $L_s^3$ : | None | r | r |
| $L_s^4$ : | 0 | 25 | 40 |
| $L_s^5$ : | 0 | 50% | 0 |

(a)          (b)

Figure 6. Example of solution representation for PV2VC problem: (a) ER, (b) ES

### 4.3.2  Initial population, evaluation and selection

The EVRP results are adopted as the initial population for solving PV2VC problem. As for evaluation and selection, we rank and select the feasible solutions with lowest ER costs. If a solution is infeasible, we again record the energy shortage of ER as the penalty and the infeasible solutions with lowest ER energy shortage are selected with higher priority. Both feasible and infeasible results are passed to the next generation. At the end of last generation, the PV2VC routes and assignments with the most cost savings are selected as the output. If there is no feasible PV2VC solution, the EVRP results are returned as the final solution.

### 4.3.3  Genetic operators

The same set of charging station operators and dwell time operators from EVRP are applied in this section. The platoon operators in Section 4.2.3 are also adopted to find the platoons between ER and ES for the PV2VC assignment. Similar to the dwell time operators in Section 4.1.3, a series of PV2VC time operators are developed here which is based on the platoon assignment results between ER and ES.

- *PV2VC full charge*: Select a random PV2VC segment and set the associated percentage of charging time to 100%. Update associated $L_s^5$. Skip if there is no ER-ES platoon.
- *PV2VC empty charge*: Select a random PV2VC segment and set the associated percentage of charging time to 0%. Update associated $L_k^3$ and $L_s^5$. Skip if there is no ER-ES platoon.
- *PV2VC random charging time (feasible)*: If current assignment is feasible, select a random PV2VC segment and randomly generate a new percentage of charging time between 0 to current value. Update associated $L_s^5$. Skip if there is no ER-ES platoon.
- *PV2VC random charging time (infeasible)*: If current assignment is infeasible, select a random PV2VC segment and randomly generate a new percentage of charging time between current value to 100%. Update associated $L_s^5$. Skip if there is no ER-ES platoon.

## 5.  Computational experiments

Numerical experiments are conducted to (1) evaluate the computational performance of the MILP model and the proposed algorithm, and (2) explore the potential benefits of platoon-based vehicle-to-vehicle charging technology. The proposed MILP model is solved using a benchmark commercial solver, Gurobi 10.0.1, with default parameter settings (runtime limit set to 2 hours). Three operation scenarios are tested and compared in this section: (1) the electric vehicle routing problem (EVRP), where vehicles can only get charged at charging station (CS), (2) the electric vehicle platooning problem (EVPP), which is an extension to the EVRP but vehicles are allowed to travel in platoon to save energy consumption, (3) the platoon-based vehicle-to-vehicle charging (PV2VC) problem, where electricity requests (ERs) can be charged by electricity suppliers (ESs) while moving to save the detour and charging time at CS. All the numerical experiments are conducted on a MacBook Air with Apple M2 CPU and 16 gigabyte RAM.

## 5.1 Experiment settings

We use the general structure of the Sioux Falls network provided by Stabler (2023) but with scaled up link costs (changed from minutes to units of miles) to accelerate energy consumption for our tests. The Sioux Falls network with the modified travel distances (miles) is shown in **Figure 7**.

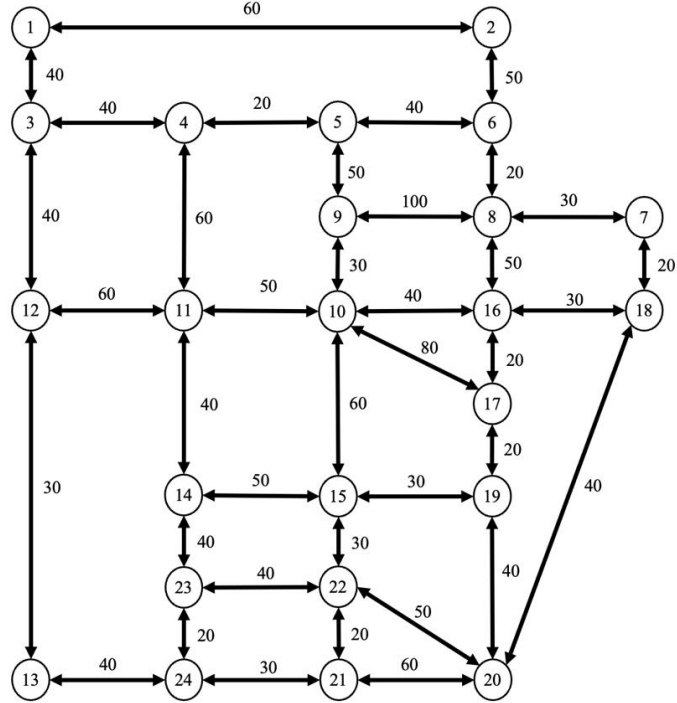

Figure 7. Sioux Falls network with modified link costs as scaled up distances

Five scenarios ($S$) are tested on the modified Sioux Falls network. Each scenario has a different number of ERs and ESs: ($S1$) $|R| = 2, |S| = 1$, ($S2$) $|R| = 3, |S| = 1$, ($S3$) $|R| = 3, |S| = 2$, ($S4$) $|R| = 4, |S| = 1$, and ($S5$) $|R| = 4, |S| = 2$. Each scenario is used to generate three instances, resulting in a total of 15 instances that are solved. Given that the GA is a stochastic search algorithm, each of the instances is solved twice and averaged. The results report the average of the six runs (three instances, two runs each) per scenario for the GA, and average of the three instances for the commercial solver.

The parameters used in the numerical experiments are summarized in **Table 4**. For convenience, we assume that all vehicles travel at a constant speed of 60 miles per hour. The charging rate at a CS is set to 180 kW, whereas the PV2VC output rate is assumed to be 50 kW. The energy consumption rate is set to 0.4 kWh/mile for all links. For those segments where vehicles travel in a platoon, they receive a constant 10% reduction in energy consumption as the platoon saving rate. For the PV2VC process, the ER can only receive 90% of the power transferred from the ES (V2V power transfer efficiency). The battery capacities of ER and ES are set to 100 kWh and 200 kWh respectively. The minimum state of charge at any location is required to be not less than 2 kWh for an ER. The inputs for all the 15 instances, including ER routes, ES initial locations, state of charge, and charging station locations, are available on Github (Fu, 2025).

Table 4. Parameters and values in numerical experiments

| Parameters | Values |
| --- | --- |
| Vehicle travel speed | 60 miles per hour |
| Charging rate at CS | 180 kW per hour |
| PV2VC charging rate | 50 kW per hour |
| PV2VC power transfer efficiency | 90% |

| Energy consumption rate | 0.4 kWh per mile |
|---|---|
| Platoon energy consumption saving rate | 10% |
| ER battery capacity | 100 kWh |
| ES battery capacity | 200 kWh |
| ER minimum required state of charge | 2 kWh |

## 5.2 Optimality gap
### 5.2.1 Summary
The comparison between the 3-run averages of the commercial solver under 2-hour runtime and the 6-run averages of the proposed GA is shown in **Table 5** and **Table 6. Table 5** summarizes the total cost objective value gap, whereas **Table 6** presents the energy consumption cost and travel time cost gaps individually. The optimality gap is measured by $\frac{(GA-MILP)}{MILP} \times 100\%$. According to our tests, the MILP model solved with the benchmark commercial solver using branch-and-bound/-cut methods cannot even handle the scenario $S5$ within the 2-hour computation time limit. Meanwhile, the proposed GA can generate a better solution within a fraction of the time. In this case, a negative optimality gap value is used and indicates that the proposed GA outperforms the commercial software within the runtime threshold.

Among the three scenarios, the proposed GA performs best in the EVRP scenario with only a 0.12% gap in terms of total cost. The average optimality gap is 0.48% in the EVPP scenario. As for the PV2VC scenario, the optimality gap increases to 1.26% among experiments S1 to S4. In S5, the proposed GA can reach solutions with 2.94% less cost than the commercial solver which times out at two hours. Overall, the proposed GA only has 0.34% optimality gap over all scenarios compared with the commercial solver.

Table 5. Objective value gap of GA under 3 instances and 2 runs each relative to the commercial solver for the 3 instances with 2-hour runtime

| Scenarios | S1 | S2 | S3 | S4 | S5 |
|---|---|---|---|---|---|
| **EVRP** | 0% | 0.03% | 0.1% | 0.21% | 0.26% |
| **EVPP** | 0.19% | 0.28% | 0.26% | 0.75% | 0.9% |
| **PV2VC** | 0.34% | 0.57% | 1.14% | 2.98% | -2.94%* |

Note: * indicates that GA outperforms the commercial solver within the 2-hour limit.

Table 6. Energy consumption cost and travel time gap breakdown of the results from Table 5

| | Energy consumption cost | | | | | Travel time | | | | |
|---|---|---|---|---|---|---|---|---|---|---|
| **Scenarios** | S1 | S2 | S3 | S4 | S5 | S1 | S2 | S3 | S4 | S5 |
| **EVRP** | 0.00% | 0.00% | 0.00% | 0.07% | 0.34% | 0.00% | 0.04% | 0.14% | 0.27% | 0.22% |
| **EVPP** | 0.03% | 0.77% | 0.25% | 0.68% | 1.33% | 0.25% | 0.10% | 0.26% | 0.77% | 0.75% |
| **PV2VC** | 0.22% | 1.00% | 1.23% | 2.51% | -2.15%* | 0.39% | 0.41% | 1.10% | 3.15% | -3.22%* |

Note: * indicates that GA outperforms the MILP within limited hours.

### 5.2.2 Breakdown of one sample run
We present more details of one sample run for PV2VC S3 with the initial parameters on Sioux Falls network as shown in Figure *8*.(a). Three ERs are given routes of 1→13→20, 4→12→13→22, and 2→5→15→22. For example, the route of $ER1$ is denoted as $ER1^1$ for node 1, $ER1^2$ for node 13, and $ER1^3$ for node 20, representing the visit sequence for $ER1$. Their initial state of charge for ERs is [20,20,25]. Three charging stations are located at nodes [3,6,20]. Given two ESs with full initial state of charge, they are located at CSs [3,6] respectively.

The optimal solution from commercial solver Gurobi is shown in Figure *8*.(b). For $ER1$ and $ER2$, they first travel to CS at node 3 and spend 4.13 mins for charging. This is due to the minimum requirement of battery level for ER at any location, which requires $ER1$ and $ER2$ to have at least 2kWh when they arrive the next node. Then they both travel in platoon with $ES1$ along the path 3→12→13→24→21, whereas only $ER1$ resumes with $ES1$ for the segment 21→20. Along the ER-ES platoon path, $ES1$ provides PV2VC service to $ER1$ with 100% of time while traversing the segment 3→12. However, for the segment 12→13, only 62.2% of time is used to provide PV2VC service to $ER2$. After arriving node 21, $ER2$ moves to its destination by itself while $ER1$ travels with $ES1$ in platoon to node 20, where 22% of the time receiving PV2VC service. As for $ER3$ and $ES2$, they start traveling in platoon at node 6 while providing PV2VC service until node 10 and then move together to the final destination of $ER3$. Although there is charging station at node 6, $ER3$ did not stop or receive any charge from the CS. Instead, it continues the trip seamlessly and gets charged while moving to the destination with $ES2$. In summary, the energy consumption costs for ERs are 88 kWh, 74.4 kWh, and 95.6 kWh, and the total energy cost is 258 kWh. The travel times for ERs are 244.13 mins, 204.13 mins, and 260 mins, which sum to 708.26 mins. The total cost then equals 966.26 mins (assuming 1 kWh equivalent in cost to 1 min).

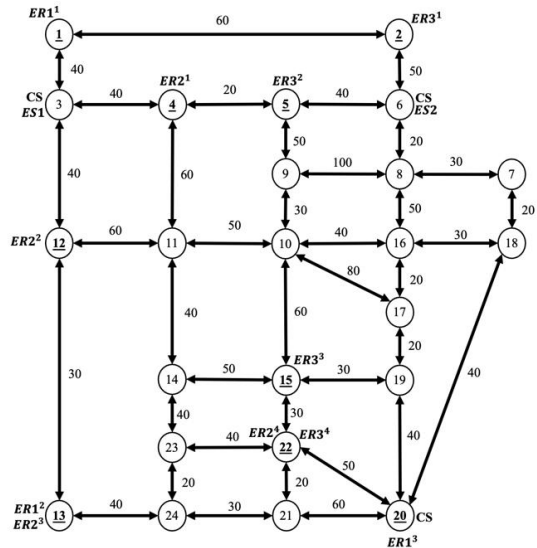

(a) Initial parameters on Sioux Falls network

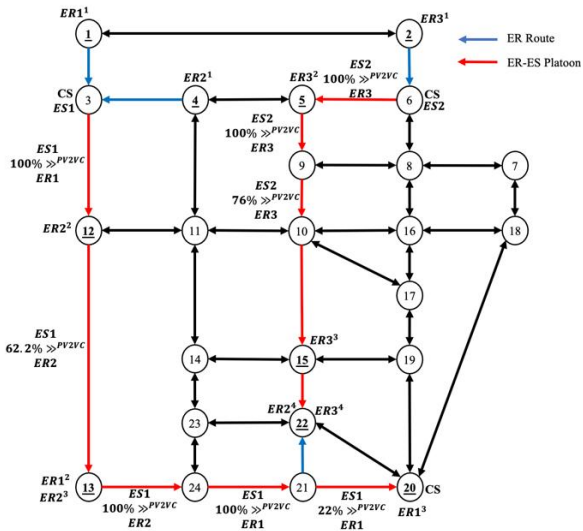

(b) Optimal solution from commercial solver Gurobi

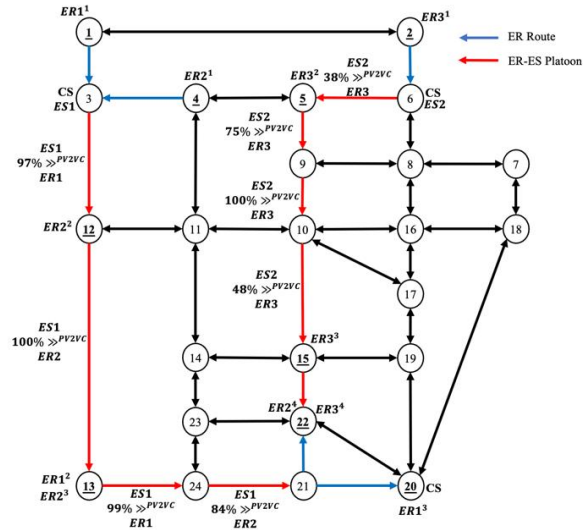

(c) Sample solution from the GA

Figure 8. Breakdown of one sample run for PV2VC S3

As for the GA, one example solution is shown in Figure 8.(c). $ER1$ and $ER2$ meet at CS at node 3 but spend 5.4 mins here for charging. Then they travel in platoon together with $ES1$ along the path 3→12→13→24→21, whereas $ER1$ and $ER2$ split at node 21 and continue to their destinations individually this time. Along the ER-ES platoon path, $ES1$ provides PV2VC service to $ER1$ with 97% of time while traversing the segment 3→12. For the segment 12→13, 100% of the time is used to provide PV2VC service to $ER2$. Next, $ES1$ spends 99% of the time of segment 13→24 to charge $ER1$ and 84% of the time of segment 24→21 to charge $ER2$. As for $ER3$ and $ES2$, they still travel in platoon along the same path, 6→5→9→10→15→22, as the commercial solver example. However, the percentage of time for PV2VC service is different across segments in this case. Overall, the energy consumption costs for ERs are 90.4 kWh, 74.4 kWh, and 95.6 kWh, and the total energy cost is 260.4 kWh. The travel times for ERs are 245.4 mins, 205.4 mins, and 260 mins, which sums to 710.8 mins. The total cost then equals 971.2 mins. As a result, the optimality gap against the optimal commercial solver solution is 0.93% in energy consumption, 0.36% in travel time, and 0.51% in total cost.

## 5.3 Computation time

The computation time is summarized in **Table 7**. The MILP model solved by the benchmark commercial solver Gurobi cannot even handle the scenario S5 within the 2-hour computation time limit, showing a gap against a lower bound of 8.84% when terminated. The required computation time of the commercial solver increases significantly with the number of ERs and ESs, especially for the PV2VC scenario, whereas the proposed genetic algorithm only requires a fraction of the time and outperforms it in large cases.

Table 7. Computation time (sec) over various scenarios

| Method | Scenarios | S1 | S2 | S3 | S4 | S5 |
|---|---|---|---|---|---|---|
| Commercial solver | EVRP | 3.6 | 3.8 | 7.1 | 13.4 | 14.9 |
| | EVPP | 4.5 | 17.2 | 37 | 61.3 | 68.1 |
| | PV2VC | 11.7 | 260.5 | 1744.5 | 5633.2 | 7200* |
| GA | EVRP | 0.4 | 1.5 | 4.3 | 5.8 | 6.7 |
| | EVPP | 1.4 | 5.7 | 15.7 | 15.1 | 24.9 |
| | PV2VC | 7.4 | 35.7 | 128.5 | 213.9 | 470.4 |

Note: 7200* indicates the MILP solver has reached the 2-hr runtime limit

## 5.4 Potential benefits

To evaluate the potential benefits of EVPP and PV2VC operation scenarios, we use the optimal EVRP solution from the commercial solver as the benchmark ($EVRP^*$). The total cost savings from the two other operation scenarios relative to EVRP are shown in **Table 8**, calculated as $\frac{EVRP^* - (EVPP \text{ or } PV2VC)}{EVRP^*} \times 100\%$. Moreover, energy consumption and travel time savings for each scenario are shown in **Figure 9**.

In the EVPP case, the overall cost savings can reach between 0.31% to 3.65% of the commercial software solutions whereas the proposed GA can get total savings between 0.05% and 2.92%. Although vehicles can travel in a platoon and receive energy consumption savings, they may also have different arrival times at the platoon join locations in most cases and thus add extra wait time as a penalty. It is rare to have large total cost savings in EVPP unless all the vehicles share the same vehicle routes, start times, and state of charge. From **Figure 9**, we find that the maximum energy consumption savings can reach 8.3% of the commercial software solution and 7.67% from GA in $S3$. However, the maximum travel time savings can only get up to 2.05% for the commercial software solution and 1.31% for the GA solution in $S4$. In $S3$, the final solutions even end up with more required travel times than the EVRP as a trade-off to the energy savings from vehicle platooning. Overall, only applying the platoon capability to EVs at stationary CS cannot obtain a significant cost reduction.

In the PV2VC scenario, the total cost savings increase from 3.45% to 9.04% with the commercial software solutions and from 3.12% to 11.26% with the proposed GA. On average, most cost savings still come from the energy consumption side, with 7.37% for the commercial software solution and 6.87% for the GA solution. Unlike the EVPP scenario, as shown in **Figure 9**, vehicle travel time savings are much closer to the energy consumption savings on average, which end up with 6.12% for the commercial software solution and 5.8% for the GA solution. Moreover, the most savings from all scenarios are obtained from the travel time savings from the proposed GA in S5 with up to 11.65%. The maximum energy consumption savings are achieved in S4 with 11.07%, obtained from the commercial software solution.

Table 8. Total cost savings relative to the EVRP commercial software solution

| Method | Scenarios | S1 | S2 | S3 | S4 | S5 |
|---|---|---|---|---|---|---|
| Commercial solver | EVPP | 1.06% | 1.12% | 0.31% | 3.65% | 3.14% |
|  | PV2VC | 3.45% | 4.35% | 7.02% | 9.04% | 8.48%* |
| GA | EVPP | 0.88% | 0.84% | 0.05% | 2.92% | 2.27% |
|  | PV2VC | 3.12% | 3.81% | 5.96% | 6.38% | 11.26% |

Note: * indicates the MILP solver has reached the 2-hr runtime limit.

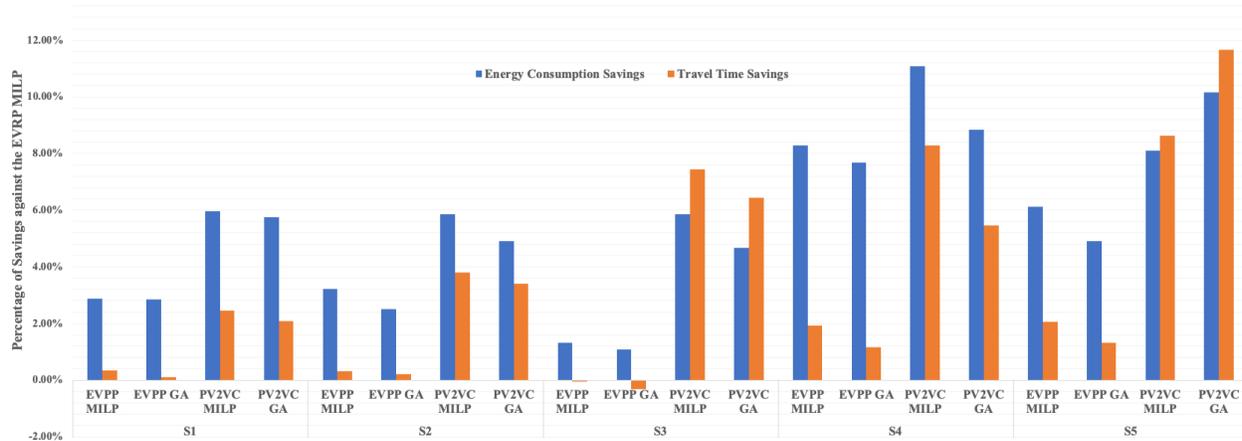

Figure 9. Energy consumption and travel time savings relative to the EVRP commercial software solution

To sum up, the EVPP operation scenario does not exhibit much potential from the EVRP case overall. The synchronization between platooned vehicles produces extra wait time and adds to their total travel time. To maximize the potential benefits from platooning, vehicles need to share most of their routes and have less time gap at charging stations or platoon join locations, which is not common in most real-world instances. However, if given a large fleet of vehicles in the same area, it would create more opportunities and make it easier to find vehicle platoons that share similar origins-destinations and travel times. As for the PV2VC, it can achieve more than 11% of total cost savings from our numerical experiments by comparison with the basic EVRP scenario. Among the five experiment scenarios, we find that the maximum savings happen to instances where the ERs start with low state of charge and are assigned with longer trips after leaving their last charging stations. In these cases, ERs can continue their routes without dwelling at the charging station and thus receive more deduction in their total travel time. The PV2VC operation scenario is even more beneficial when we consider the queueing time and detours at charging stations. Moreover, in cases where the travel time is perceived much more valuable than the energy cost, such as medical emergency or disaster evacuation, PV2VC can minimize the travel time for ERs as much as possible.

## 6. Conclusion

With the development of autonomous and connected vehicle technology and the promotion of electric vehicles, we introduce an innovative solution to mitigate the insufficient charging facility problem for the charging process of EVs. This new dynamic charging solution, defined as the PV2VC technology, is able to dispatch a fleet of ESs to charge the ERs while moving together in a platoon. Compared with stationary charging solutions such as charging stations, mobile chargers (CaaS), and battery swapping stations (BaaS), EVs can avoid the detour to charging stations and still be able to move towards the destination without stopping for charging.

To evaluate the operational viability of PV2VC technology, we formulate a MILP model to jointly optimize the routing and charging scheduling of ERs and ESs to minimize the energy consumption and travel time for ER. If an ER travels in platoon with other vehicles, the energy consumption cost receives a constant reduction along the platoon segments. Since ERs and ESs need to travel in proximity in a platoon to transfer electricity between vehicles, we track the movement and status of EVs and ensure the temporal-spatial synchronization for the precision and efficiency of the entire operation, especially the platoon formation and separation process. In addition, we consider two more basic operation scenarios, the electric vehicle routing problem (EVRP) and the electric vehicle platooning problem (EVPP). EVs can only get charged at stationary locations (partial charge allowed) and travel independently in the benchmark scenario (EVRP), whereas vehicles are allowed to wait and form platoons along the path to reduce their energy consumption in EVPP. Both the EVRP and EVPP operation scenarios are mathematically formulated in the MILP model as well, with minor modifications from the PV2VC problem.

We propose a GA for large-scale instances of EVRP, EVPP, and PV2VC problems. For the EVRP, each electricity request is considered as an independent sub-problem. We develop a two-layer structure (visiting sequence and dwell time) to represent the solution, a heuristic algorithm to construct the initial population, and two sets of genetic operators to modify the locations and charging time at CSs for each ER. In the EVPP, two more layers (platoon vehicles and arrival time along the path) are added into the solution representation and one more set of platoon-based genetic operators are proposed to search and identify electric vehicle platoons. As for the last scenario (PV2VC), the previous solution representations are slightly modified to distinguish ERs and ESs. One more layer is introduced specifically for the ES to denote the PV2VC percentage along the path. The platoon-based genetic operators are adopted from the EVPP and integrated with a new set of PV2VC time operators to find ER-ES platoon assignments.

To evaluate the computational performance of our proposed genetic algorithm and explore the potential benefits of the PV2VC technology, numerical experiments are implemented on the Sioux Falls network with modified link costs. Five scenarios are tested, each with three different instances involving different numbers of ERs and ESs. The proposed GA is compared with the MILP model solved by a benchmark commercial solver Gurobi within a 2-hour runtime limitation. The results show that the proposed GA can reach an average optimality gap of 0.34% against the MILP model with only a fraction of required computation time. By comparing against the EVRP solutions obtained from the MILP model, the PV2VC can save up to 11.26% in total cost, 11.07% in energy consumption and 11.65% in travel time, whereas the EVPP can save up to 3.65% savings in total cost, 8.3% in energy consumption and 2.05% in travel time instead.

For future research, larger scale and more diverse real-world settings can be tested. Second, the MILP model and proposed GA can be extended for dynamic and online applications to simulate daily operations. Third, electric vehicle relocation problem can be considered and combined with the routing and scheduling of ER and ES fleets. Lastly, the ES can be integrated with other practical applications such as deliveries when not being used for the PV2VC service.

**Acknowledgments**

This research was partially supported by the C2SMARTER Center, under USDOT #69A3551747124.